\documentclass[%
 aip,
 amsmath,amssymb,
 reprint,%
]{revtex4-1}

\usepackage{graphicx}
\usepackage{dcolumn}
\usepackage{bm}

\usepackage[utf8]{inputenc}
\usepackage[T1]{fontenc}
\usepackage{mathptmx}
\usepackage{etoolbox}

\makeatletter
\def\@email#1#2{%
 \endgroup
 \patchcmd{\titleblock@produce}
  {\frontmatter@RRAPformat}
  {\frontmatter@RRAPformat{\produce@RRAP{*#1\href{mailto:#2}{#2}}}\frontmatter@RRAPformat}
  {}{}
}%
\makeatother
\begin{document}

\preprint{AIP/123-QED}

\title[Compact Fully Integrated..]{Compact Fully Integrated Alignment-Free and Linearly Polarized Ho$^{3+}$ Single-Frequency Waveguide Laser}
\author{Jeswin Jacob}
\affiliation{ 
School of Physics, Chemistry, and Earth Sciences, College of Science, Adelaide University, Mawson Lakes Campus, South Australia - 5095
}%
\author{T Toney Fernandez$^{*}$}
\affiliation{ 
School of Physics, Chemistry, and Earth Sciences, College of Science, Adelaide University, Mawson Lakes Campus, South Australia - 5095
}%
\email{toney.teddyfernandez@adelaide.edu.au}
\author{Dale E. Otten}
\affiliation{ 
School of Physics, Chemistry, and Earth Sciences, College of Science, Adelaide University, Mawson Lakes Campus, South Australia - 5095
}%
\author{Ji Eun Bae}
\affiliation{ Centre de Recherche sur les Ions, les Matériaux et la Photonique (CIMAP), UMR 6252 CEA-CNRS-ENSICAEN, Université de Caen Normandie, 6 Boulevard Maréchal Juin, 14050 Caen, France
}
\author{Ori Henderson-Sapir}
\affiliation{ 
Institute for Photonics, Advanced Sensing and Quantum Technologies (IPAS-QT) and the School of Physics, Chemistry and Earth Sciences, Adelaide University, 5005, SA, Australia
}
\affiliation{ 
OzGrav - Australian Research Council Centre of Excellence for Gravitational Wave Discovery, Adelaide Node, SA, Australia
}
\author{Pavel A. Loiko}
\affiliation{ Centre de Recherche sur les Ions, les Matériaux et la Photonique (CIMAP), UMR 6252 CEA-CNRS-ENSICAEN, Université de Caen Normandie, 6 Boulevard Maréchal Juin, 14050 Caen, France
}

\author{Karen Privat}
\affiliation{ Electron Microscope Unit, Mark Wainwright Analytical Centre, University of New South Wales, Sydney, Australia}

\author{David Lancaster}
\affiliation{ 
School of Physics, Chemistry, and Earth Sciences, College of Science, Adelaide University, Mawson Lakes Campus, South Australia - 5095
}%
\date{\today}

\begin{abstract}
We report, for the first time, a fully integrated Ho$^{3+}$-doped fluoride glass distributed Bragg reflector (DBR) waveguide laser, in which a femtosecond laser-inscribed waveguide Bragg grating (WBG) provides wavelength-selective feedback within the laser cavity. For comparison, we also demonstrate a Ho$^{3+}$-doped fluoride glass waveguide laser incorporating an external fiber Bragg grating (FBG). A line-by-line inscription technique was employed to fabricate the Bragg gratings, yielding highly consistent and reproducible grating characteristics with precise control of the Bragg resonance wavelength. The integrated DBR laser, featuring a compact 12.4 mm cavity, achieves stable single-frequency lasing at 2062.21 nm with a linewidth of less than 5 MHz. The laser exhibits intrinsically linearly polarized emission. This alignment-free architecture provides a compact and robust platform for integrated mid-infrared single-frequency laser sources.
\end{abstract}

\maketitle

\section{\label{sec:level1}Introduction}

Compact, alignment-free single-frequency lasers have become increasingly important for applications requiring highly coherent and spectrally stable optical sources~\cite{Fu2017}. In particular, they play a key role in coherent optical communication systems, where narrow-linewidth emissions enable high-order modulation formats and improved signal integrity, and in optical frequency comb generation, where stable single-frequency pump lasers are essential for low-noise comb formation and precise frequency referencing~\cite{Poulsen2003, Pavlov2018}. Concurrently, the growing demand for portable and robust photonic systems has driven the development of integrated laser architectures that eliminate complex free-space alignment while maintaining excellent spectral performance~\cite{Bernier:15,Hudson:13,Guay:19}. Femtosecond laser-written waveguide lasers offer a promising route toward such compact sources by combining micrometer-scale three-dimensional fabrication flexibility with low-loss light confinement~\cite{Marshall08, Ams:09}. However, fully integrated single-frequency waveguide lasers operating in the 2 $\mu$m spectral region remain scarce to non-existent~\cite{Lily2019, Scholle10}. This scarcity has been particularly pronounced in glass-based photonic systems, where the availability of suitable low-loss waveguides in the 2~$\mu$m spectral region has historically been limited. Early femtosecond-laser-written structures in the infrared were predominantly based on depressed-cladding geometries~\cite{Lancaster, Gross12} or low-contrast positive-index changes~\cite{Ledemi14, Berube13}, which provided limited mode confinement and restricted the development of compact, fully integrated laser architectures. The recent emergence of high positive-index-contrast waveguides in infrared-transmitting glasses has opened new opportunities for stronger confinement and the integration of fiber laser components within a single monolithic photonic platform~\cite{FernandezOE24,An}. Here, we report the first fully integrated, alignment-free Ho$^{3+}$-doped fluoride glass DBR single-frequency waveguide laser, employing a femtosecond laser-inscribed waveguide Bragg grating chip bonded to a high-reflectivity mirror. The resulting compact cavity achieves stable single-frequency lasing at 2062.21 nm.

\section{\label{sec:level1}Experimental Methods}

\begin{figure*}
\includegraphics[trim={0 0 0 0},width=12 cm]{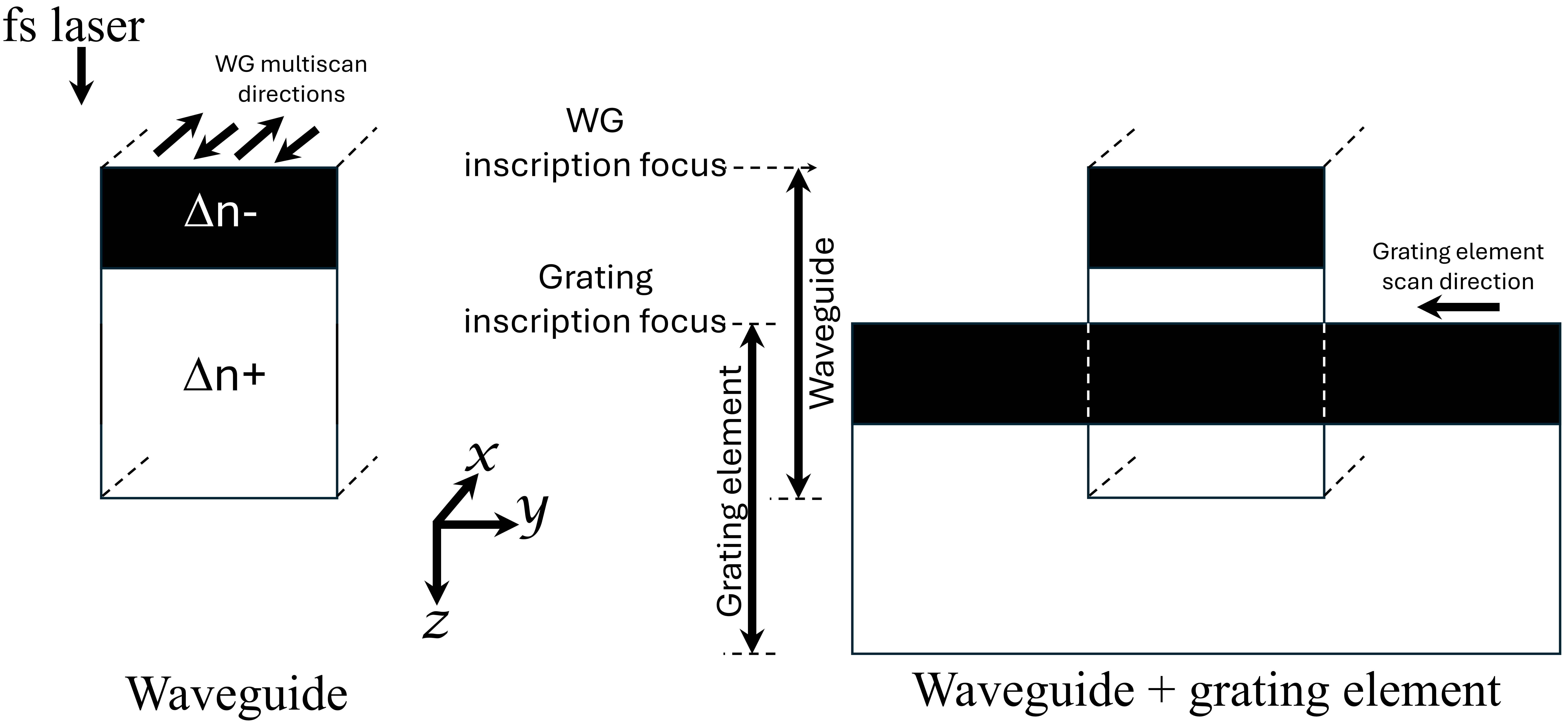}
\caption{Waveguide and waveguide~+~grating plane cartoon. The black solid fill represents the negative index change and the white rectangles represent the positive index change induced by the fs-laser. In all cases the inscribing femtosecond laser was incident from the top of the page. The waveguides were fabricated by focusing the laser $170\,\mu\mathrm{m}$ and the grating planes between $182-188\,\mu\mathrm{m}$ below the top surface. The inscription direction of the waveguide (x-axis) and grating planes (y-axis) were orthogonal to each other as shown.
\label{Cartoon}}
\end{figure*}

The Ho$^{3+}$-doped fluoride glass used in this work had the tailored composition of $53 ZrF_4\-~22 BaF_2\-~3AlF_3\-~20NaF\-~2HoF_3(mol\%)$. This composition was selected based on our previous studies, which demonstrated its ability to support high refractive index contrast waveguides inscribed by femtosecond laser direct write process~\cite{FernandezOE24}.
Waveguides and waveguide Bragg gratings (WBGs) were inscribed using a Pharos femtosecond laser system (Light Conversion) operating at a central wavelength of 1030~nm with a pulse duration of 240~fs, laser power of 90 mW at a repetition rate of 50~kHz. The structures were fabricated at a translation speed of 40~$\mu$m/s using a 0.65 NA objective whose correction collar was set at 500~$\mu$m, which was previously identified as the optimum configuration for producing high index change waveguides in this glass. The waveguides can be readily designed to match the single-mode field of standard optical fibers at different wavelengths by simply adjusting the waveguide dimensions via the number of scans used in the multi-scan fabrication process. This versatility enables mode matching to a wide range of standard fibers, including HI1060, SM980, SMF28, SM1950 and their polarization-maintaining equivalents~\cite{fernandez2024}. At first the waveguides were inscribed at a depth of 170~$\mu$m below the sample surface, with a width of $9~\mu\mathrm{m}$. This width was selected to support single-mode operation at a wavelength of $2~\mu\mathrm{m}$ while enabling efficient coupling to and from SM1950 single-mode fiber with low insertion loss~\cite{CoherentSM1950}. The waveguide Bragg gratings (WBG) were subsequently inscribed using a line-by-line technique at depths ranging from 182 to 188~$\mu$m below the sample surface. The gratings were inscribed immediately following waveguide fabrication in an automated, single-step fs-laser inscription procedure without realigning the samples. The grating depth was tuned to ensure that the negative index change associated with the grating-plane structure overlaps optimally with the positive index change guiding region of the pre-written waveguide for maximum grating strength. A cartoon of the waveguide and waveguide integrated grating plane is shown in figure~\ref{Cartoon}.

An effective refractive index of $n_{eff} = 1.485$ was initially considered for the grating design, accounting for both the bulk refractive index and the refractive index modification introduced by the waveguide, and subsequently refined through experimental iterations to $n_{eff} = 1.484$ for the WBG inscriptions. The grating period was designed to achieve the target Bragg wavelength with a fabrication tolerance of approximately $100\pm 2$~nm. Grating lengths of $L_g =$ 5, 7.5, 8.5, 10, 13.5~mm were investigated in a 21 mm long sample.
Third-order Bragg gratings were employed because first- and second-order gratings required periods comparable to the minimum feature size achievable with the current inscription parameters and focusing objective. The longer spatial period of the third-order gratings enabled reliable fabrication while maintaining high grating quality and the desired Bragg resonance.

\section{\label{sec:level1}Results and discussion}
\subsection{\label{sec:level2}Integration of grating planes into pre-written waveguides}

\begin{figure*}
\centering
\includegraphics[trim={0 0 0 0},width=10 cm]{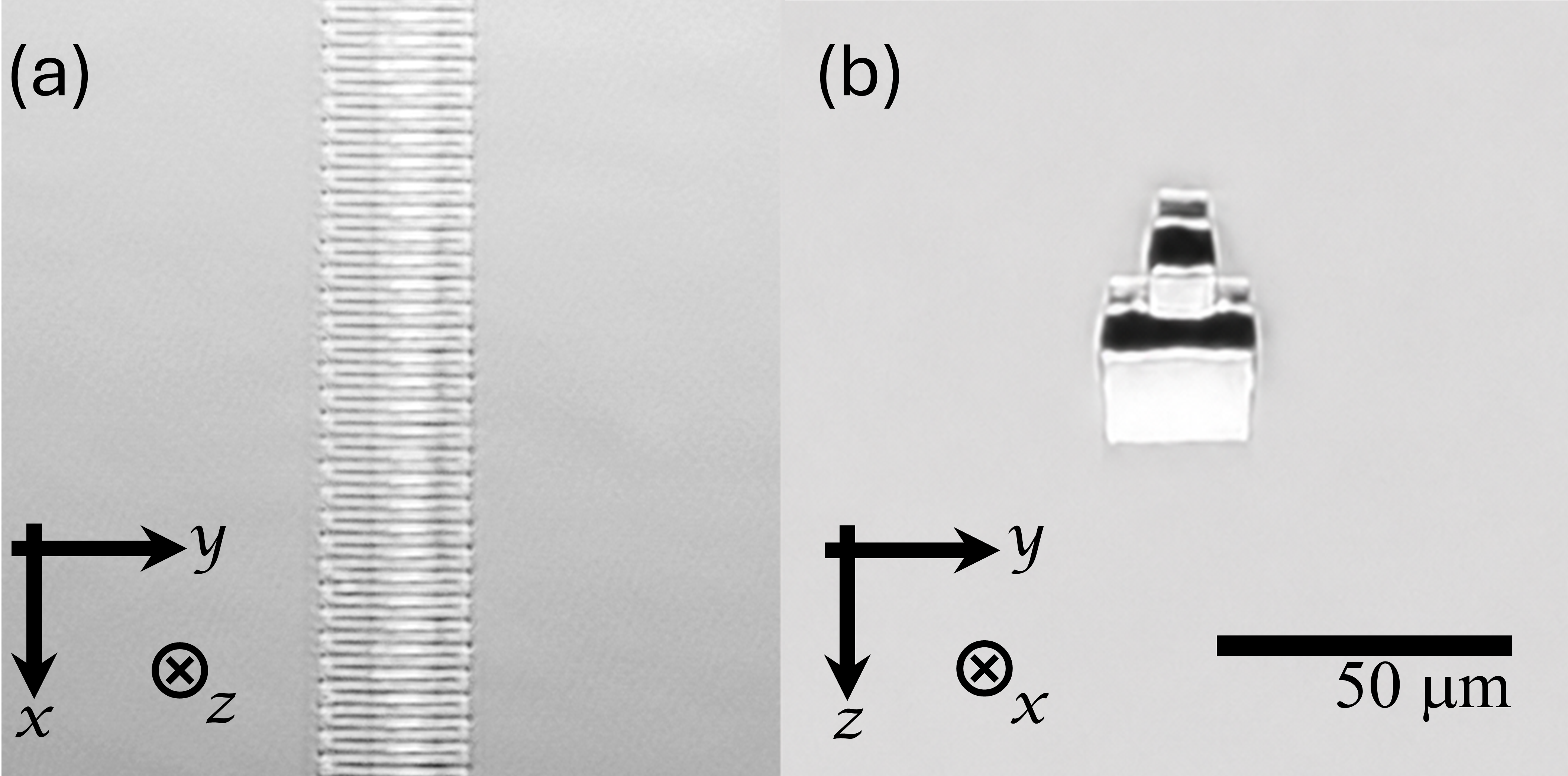}
\caption{(a) Top down view of the integrated waveguide Bragg grating inscribed in third order for 2060 nm Bragg resonance. Inscribing fs-laser is along the z-axis (b) The cross sectional view of the waveguide and the relative position of grating plane which were inscribed at depths of $170\,\mu\mathrm{m}$ and $185\,\mu\mathrm{m}$ respectively.
\label{Microscope}}
\end{figure*}

To visualize and image the relative positions of the waveguide and the grating planes, the end facets of the sample were carefully polished to expose the buried grating structures. The image shown in Figure~\ref{Microscope} illustrates one of the three different grating-plane depths with respect to the waveguide core. 

\begin{figure}
\centering
\includegraphics[trim={0 0 0 0},width=8.5 cm]{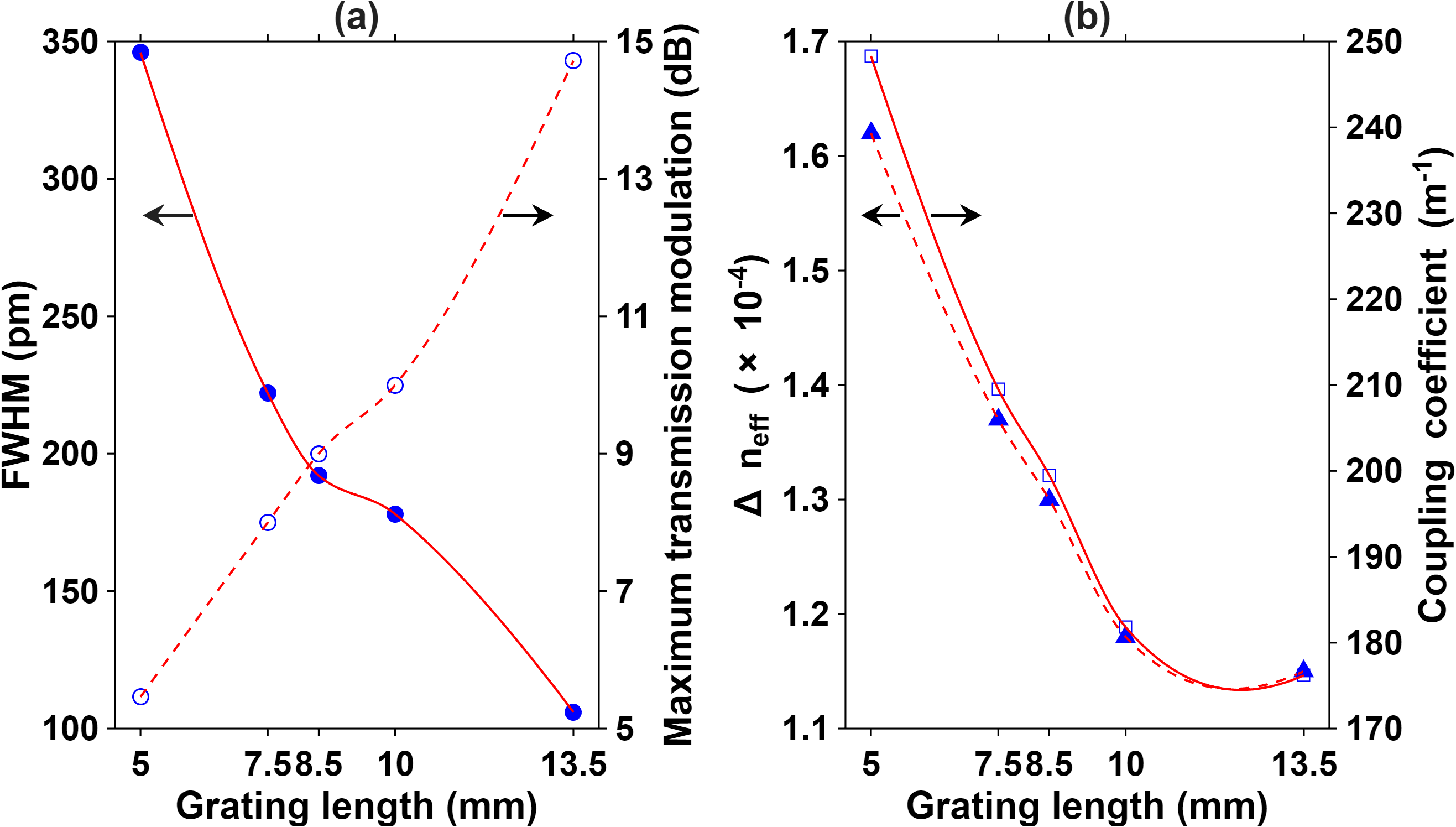}
\caption{ Variation of (a) FWHM and transmission modulation at Bragg resonant wavelength of 2055 nm along with (b) effective modal refractive index modulation and coupling coefficient.\label{Response} }
\end{figure}

Bragg gratings with resonance wavelengths spanning 2005–2080 nm were fabricated. For comparative analysis, three representative gratings with Bragg resonance wavelengths of $\lambda_B = $2055, 2060, and 2070 nm were selected.  As the gratings at these wavelengths exhibited comparable fabrication trends and spectral responses, the influence of the ultrafast laser–matter interaction on grating formation was investigated in detail using the $\lambda_B =$ 2055 nm grating as a representative case. Figure~\ref{Response} summarizes the dependence of the grating characteristics on the inscribed grating length ($L_g$), which was varied from 5 to 13.5 mm. The parameters considered include the spectral full width at half maximum ($\Delta \lambda _{\mathrm{FWHM}}$), the maximum transmission modulation at the Bragg resonance ($\lambda _{\mathrm{B}}$), the effective modal refractive index modulation of the propagating mode ($\Delta n_{eff}$), and the coupling coefficient ($\kappa$). Figure~\ref{Response}(a) demonstrates that as $L_g$ increased from 5 to 13.5 mm, the transmission modulation (dip) increased from 5.46 to 14.72 dB, corresponding to equivalent reflectivities of 71.6\% and 96.63\%, respectively. The reflectivity was estimated from the measured transmission dip as
\begin{equation}
R = 1 - 10^{-\Delta T_\mathrm{dB}/10},
\end{equation}
where $\Delta T _\mathrm{dB}$ is the transmission dip in dB, measured relative to the local out-of-band transmission level on the longer-wavelength side of the Bragg resonance. This normalization accounts for the wavelength-independent insertion loss of the waveguide device. The reflectivity estimate assumes that any additional wavelength-dependent loss introduced at the Bragg resonance is negligible.  Meanwhile, $\Delta \lambda_{\mathrm{FWHM}}$ decreased from 346 pm to an apparent value of 106 pm, indicating spectral narrowing with increasing grating length, as expected. 

Figure~\ref{Response}(b) shows that both the effective refractive-index modulation $\Delta n_{\mathrm{eff}}$, and the coupling coefficient, $\kappa$, decrease with increasing grating length. These parameters were estimated from the measured transmission dip, grating length, and Bragg wavelength using \begin{equation} \kappa = \frac{1}{L_g} \cosh^{-1}\left(10^{\Delta T_\mathrm{dB}/20}\right), \end{equation} where $L_g$ is the grating length and $\Delta T_\mathrm{dB}$ is the transmission dip in dB. The corresponding effective refractive-index modulation was calculated as \begin{equation} \Delta n_{\mathrm{eff}} = \frac{\kappa \lambda_B}{\pi}, \end{equation} where $\lambda_B$ is the measured Bragg wavelength~\cite{Kashyap2010}. Both $\kappa$ and $\Delta n_{\mathrm{eff}}$ decrease approximately linearly from $L_g =$ 5 to 10~mm, followed by a smaller variation at 13.5~mm. The observed monotonic decrease in both parameters is indicative of cumulative positioning errors originating from the Aerotech air-bearing nanopositioning stages leading to increased grating element de-phasing~\cite{Laberdesque, Feced2000}. The coupling coefficient, $\kappa$, exhibits a pronounced decrease with increasing grating length, decreasing from $248.3~\mathrm{m^{-1}}$ at $L_g =$ $5$~mm to $181.8~\mathrm{m^{-1}}$ at $10$~mm. Since $\kappa$ is derived from the measured transmission dip and grating length, the observed variation can be influenced by the repeatability of the grating inscription and sample repositioning, as well as variations in the measured grating response. Nevertheless, the relatively small change in $\kappa$ at longer grating lengths suggests that the effective coupling strength approaches a limiting value under the present inscription conditions.

\begin{figure*}
\centering
\includegraphics[trim={0 0 0 0},width=15 cm]{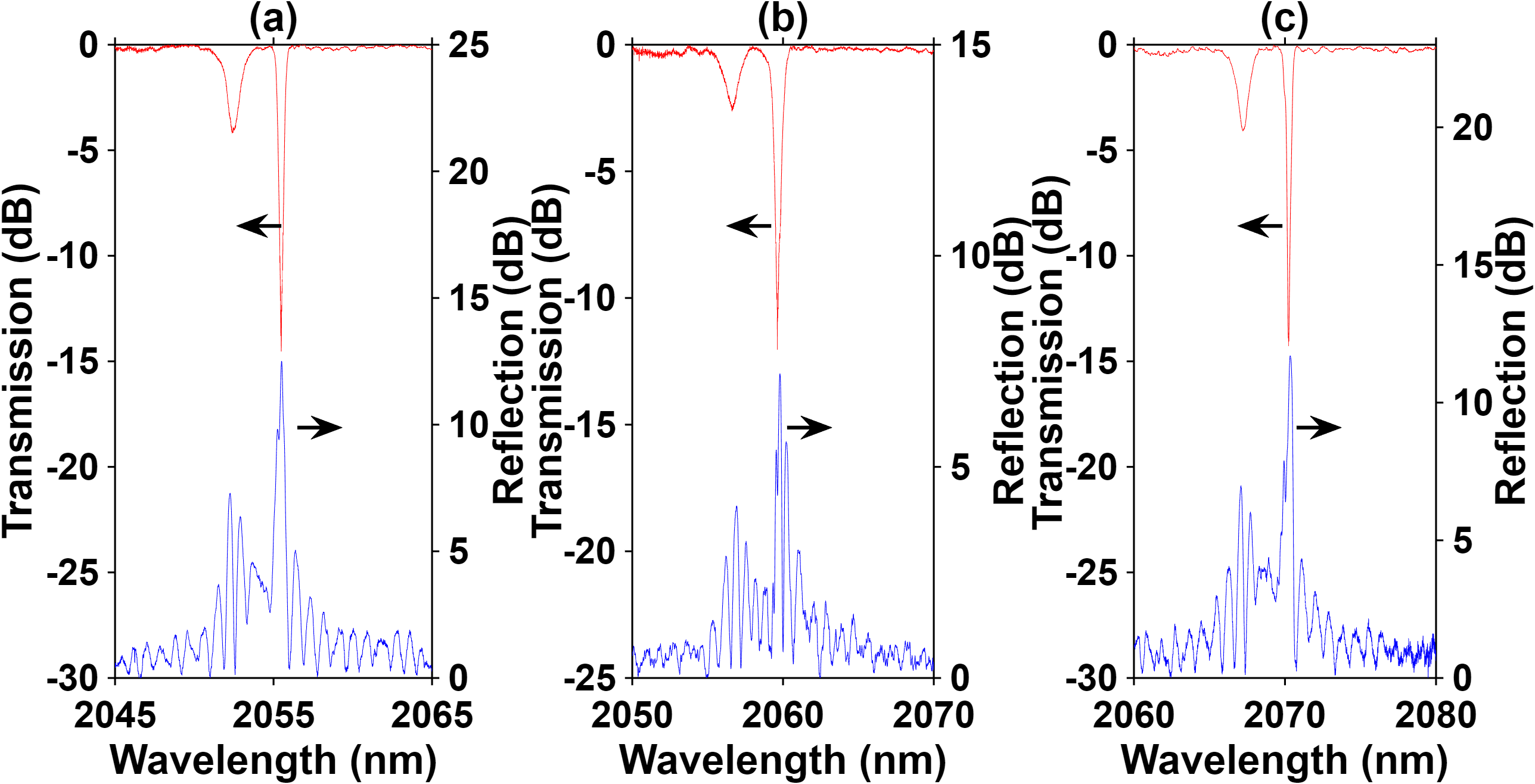}
\caption{Measured reflection and transmission spectra of 13.5-mm-long Bragg gratings designed for different resonance wavelengths.\label{Tx-Rx}}
\end{figure*}

Three Bragg grating resonances, measured in transmission and reflection at wavelengths around 2055, 2060, and 2070~nm, are presented in Fig.~\ref{Tx-Rx}. The three spectra are presented to demonstrate the repeatability and spectral tunability of the current Bragg grating inscription method. The close agreement between the Bragg resonance wavelengths observed independently in the transmission and reflection spectra confirms that both measurements correspond to the same grating resonance, consistent with coupled-mode theory~\cite{Erdogan1997}. The small variation in resonance wavelength with grating length is attributed to fabrication-induced variations in the effective grating period and/or effective refractive index.
In each spectrum, a broad, low-intensity notch is observed on the shorter-wavelength side of the narrow and pronounced Bragg resonance. This additional feature is characteristic of a cladding-like mode resonance; however, since the grating is inscribed directly within the waveguide, the precise origin and modal nature of this resonance cannot be conclusively identified at this stage. However, as per previous reports~\cite{Bhardwaj22, Fernandez24}, the low-index-change region should not be interpreted as a spatially uniform negative-index perturbation. Instead, the formation of micro- and nanovoids produces strong, spatially varying positive and negative index changes, visible in the backscatter electron microscope image shown below (Fig.~\ref{BSE}). Shorter-wavelength fields are more sensitive to these localized index features and can therefore encounter regions with an effective positive-index contrast, enabling localized confinement and the existence of radiative modes. 

\begin{figure*}
\centering
\includegraphics[trim={0 0 0 0},width=12 cm]{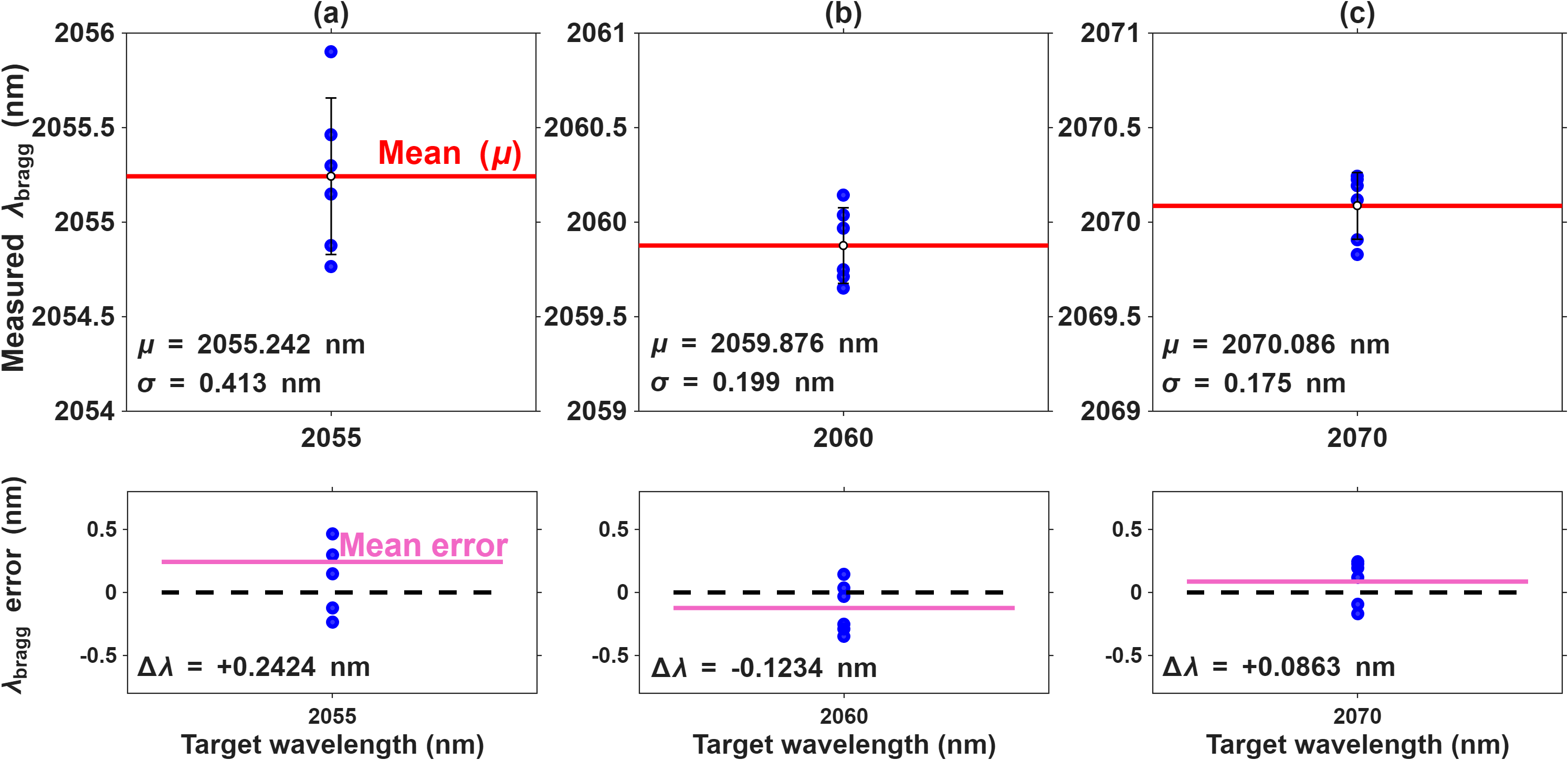}
\caption{Bragg wavelength design accuracy and repeatability of the fabricated Bragg gratings at design wavelengths of 2055, 2060, and 2070 nm. (a)–(c) Measured Bragg wavelengths for each target wavelength, showing the mean value ($\mu$) and standard deviation ($\sigma$). The corresponding wavelength deviations from the design values are shown in the lower panels.\label{Stats} }
\end{figure*}

A total of 18 Bragg gratings, designed for Bragg wavelengths of 2055, 2060, and 2070 nm, were fabricated to evaluate the accuracy of the Bragg wavelength design and the reproducibility of the resulting resonance wavelengths. The measured Bragg wavelengths were centered at 2055.2 ± 0.4 nm, 2059.8 ± 0.2 nm, and 2070.1 ± 0.1 nm, respectively. Across all 18 gratings, the measured resonances exhibited a pooled standard deviation of 280 pm. The mean wavelength variations and deviations from the target wavelengths are shown in Fig.~\ref{Stats}. The close agreement between the measured and design wavelengths demonstrates the high reproducibility and wavelength accuracy of the grating fabrication process across the investigated wavelength range.

\subsection{\label{sec:level2}Single frequency lasing cavity}
\begin{figure}
\centering
\includegraphics[trim={0 0 0 0},width=8 cm]{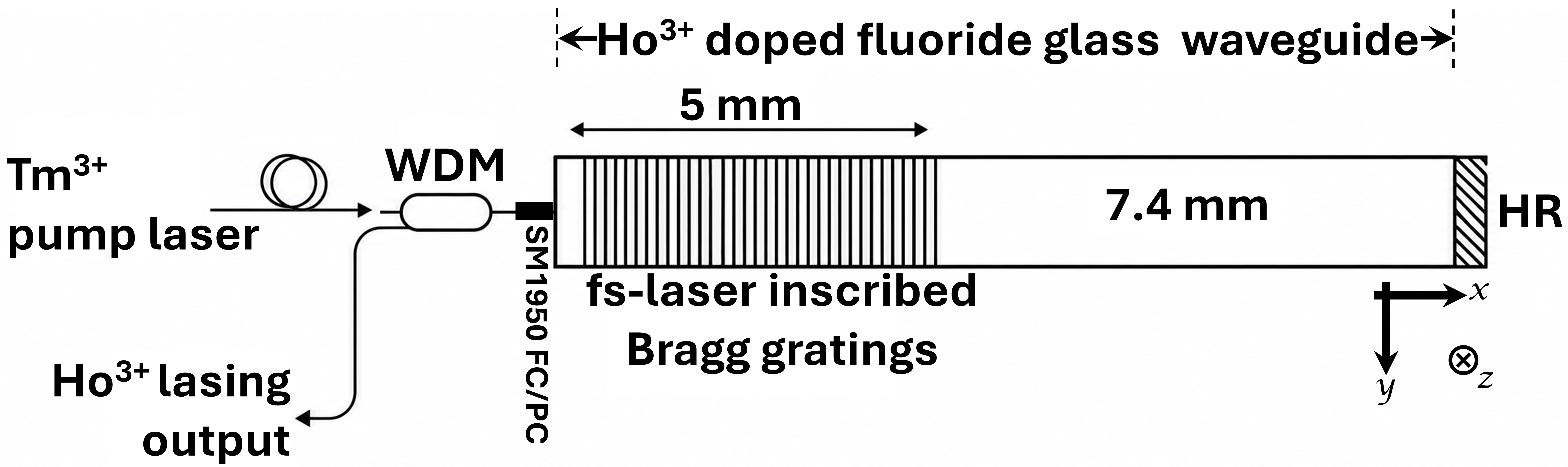}
\caption{12.4 mm long single frequency lasing cavity.\label{Lasing_Scheme} }
\end{figure}

One of the simplest laser cavities was realized using a 12.4 mm-long Ho$^{3+}$-doped fluoride glass chip. The lasing configuration is shown in Fig.~\ref{Lasing_Scheme}. (Note that this is a shorter glass sample compared to the one that was used for grating characterization.) Briefly, an SM1950 single-mode fiber was butt-coupled into the chip, enabling both waveguide pumping and simultaneous collection of the lasing output. A partially reflecting (PR) mirror was formed by a 5 mm long femtosecond (fs) laser-written Bragg grating, designed to operate at a Bragg wavelength of 2062 nm with an output coupling of 68\%. A highly reflecting (HR) mirror was bonded to the opposite facet, located roughly 7.4 mm from the end of the grating. For a uniform finite-length Bragg grating, the effective penetration depth relevant to the cavity free spectral range (FSR), considering the intensity reflection profile, can be expressed as
\begin{equation}
L_{\mathrm{p}} = \frac{\tanh(\kappa L_{\mathrm{g}})}{2\kappa},
\end{equation}
where $\kappa$ is the grating coupling coefficient and $L_{\mathrm{g}}$ is the physical grating length \cite{Barmenkov:06}. For the present 5~mm grating, with a coupling coefficient of $\kappa = 248.3~\mathrm{m}^{-1}$, the effective penetration depth is $L_{\mathrm{p}} \approx 1.7~\mathrm{mm}$. This corresponds to an effective cavity length of $L_{\mathrm{eff}} = 9.1~\mathrm{mm}$, giving a free spectral range of approximately 11.1~GHz, assuming a group refractive index of 1.484.

For lasing operation the chip was bonded to an aluminum block to enable thermal conduction. The pump source for this initial demonstration was a watt-level Tm$^{3+}$-doped fiber laser with an operating wavelength of 1945 nm, with its output delivered through an optical fiber spliced to a wavelength-division multiplexer (WDM). 
This design eliminates free-space components and enables a compact, alignment-free single-frequency laser design due to the extremely short cavity and narrow WBG resonances.

\begin{figure*}
\centering
\includegraphics[trim={0 0 0 0},width=16 cm]{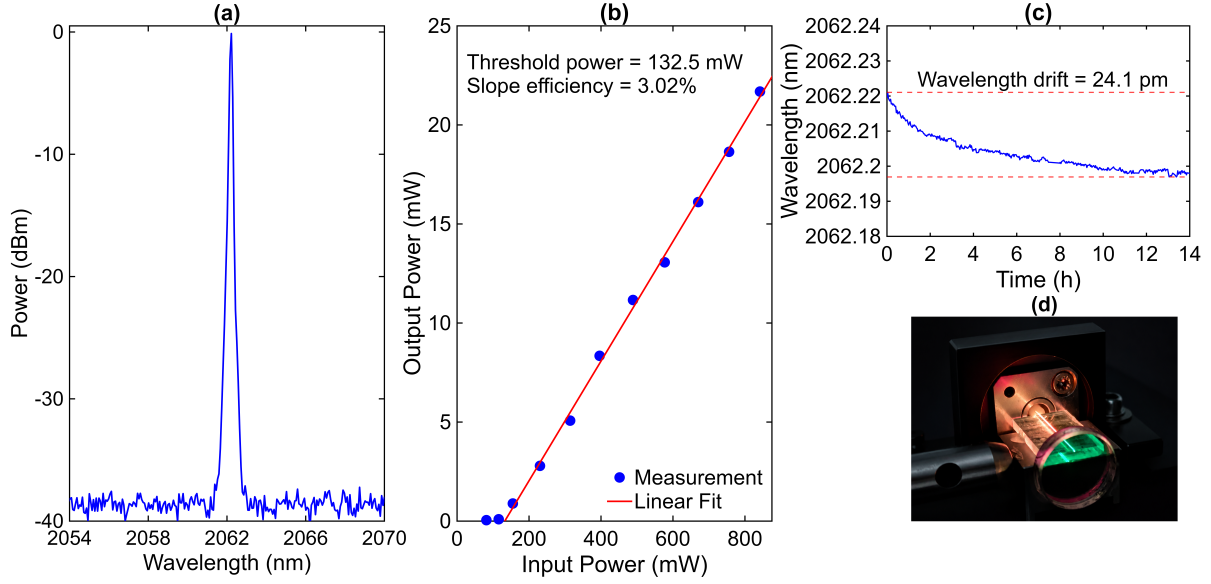}
\caption{(a) Lasing peak at 2062.21 nm (b) Lasing slope efficiency (c) Wavelength stability over 14 hours.  (d) Photograph of the compact laser cavity formed by a butt-coupled fiber ferrule and a 1/2" high-reflectivity (HR) mirror.\label{Lasing_results} }
\end{figure*}

\begin{figure*}
\centering
\includegraphics[trim={0 0 0 0 0},width=12 cm]{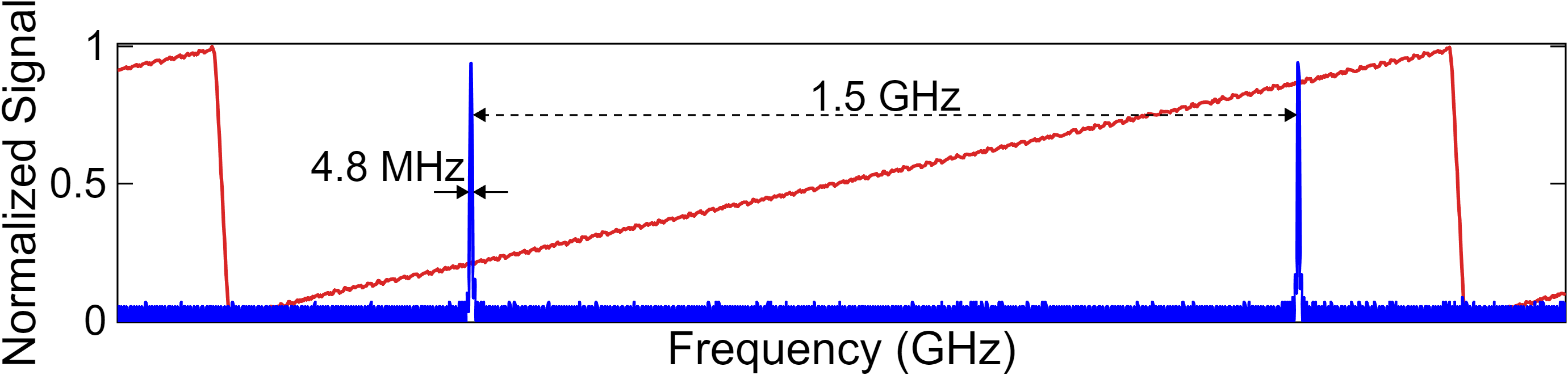}
\caption{Fabry–Pérot interferometer spectrum of the Distributed Bragg Reflector (DBR) laser output, measured with a free spectral range of 1.5 GHz. \label{FSR} }
\end{figure*}

Lasing was observed at a wavelength of 2062.21 nm (Fig.~\ref{Lasing_results}(a)), with a measured –3 dB bandwidth of 119.4 pm, which is close to the maximum resolution limit of $\sim$100 pm of the optical spectrum analyzer (Yokogawa, AQ6376E). A maximum lasing output power of 21.7 mW was achieved at an incident pump power of 825 mW, with a slope efficiency of 3.02\% and threshold pump power of 132.5 mW, as shown in Fig.~\ref{Lasing_results}(b). Single-frequency operation was verified using a scanning Fabry--Pérot (FP) interferometer with a free spectral range (FSR) of 1.5~GHz and a finesse of about 250. Figure~\ref{FSR} shows the measured FP spectrum, with a single resonance observed within the FSR, confirming single-longitudinal-mode operation. The measured linewidth of approximately 4.8~MHz is limited by the resolution of the interferometer and therefore represents an upper bound on the laser linewidth. No evidence of mode hopping or spectral instability was observed during the measurement.

The long-term wavelength stability was characterized using a laser spectrum analyzer (Bristol 771B) with a spectral resolution of 2 GHz and an absolute accuracy of 100 fm, at a pump power of 1 W. Measurements were conducted over a period exceeding 14 hours under ambient laboratory conditions, including uncontrolled laboratory temperature (e.g. no A/C), and no active thermal stabilization of the cavity. Despite these conditions, the lasing wavelength exhibited a drift of only~24 pm over the entire measurement duration (Fig.~\ref{Lasing_results}(c)). This result highlights the excellent stability of an alignment-free laser system and also demonstrates the reliability of the current line-by-line femtosecond laser inscription technique for fabricating Bragg gratings. Fig.~\ref{Lasing_results}(d) shows the compact cavity employed in this work, consisting of a butt-coupled fiber ferrule and a high-reflectivity (HR) mirror. We believe that the extremely short cavity length and minimal number of optical components actively contributed to the mechanical and thermal stability of the laser, supporting the observed wavelength stability.

\begin{figure}
\includegraphics[trim={0 0 0 0},width=8 cm]{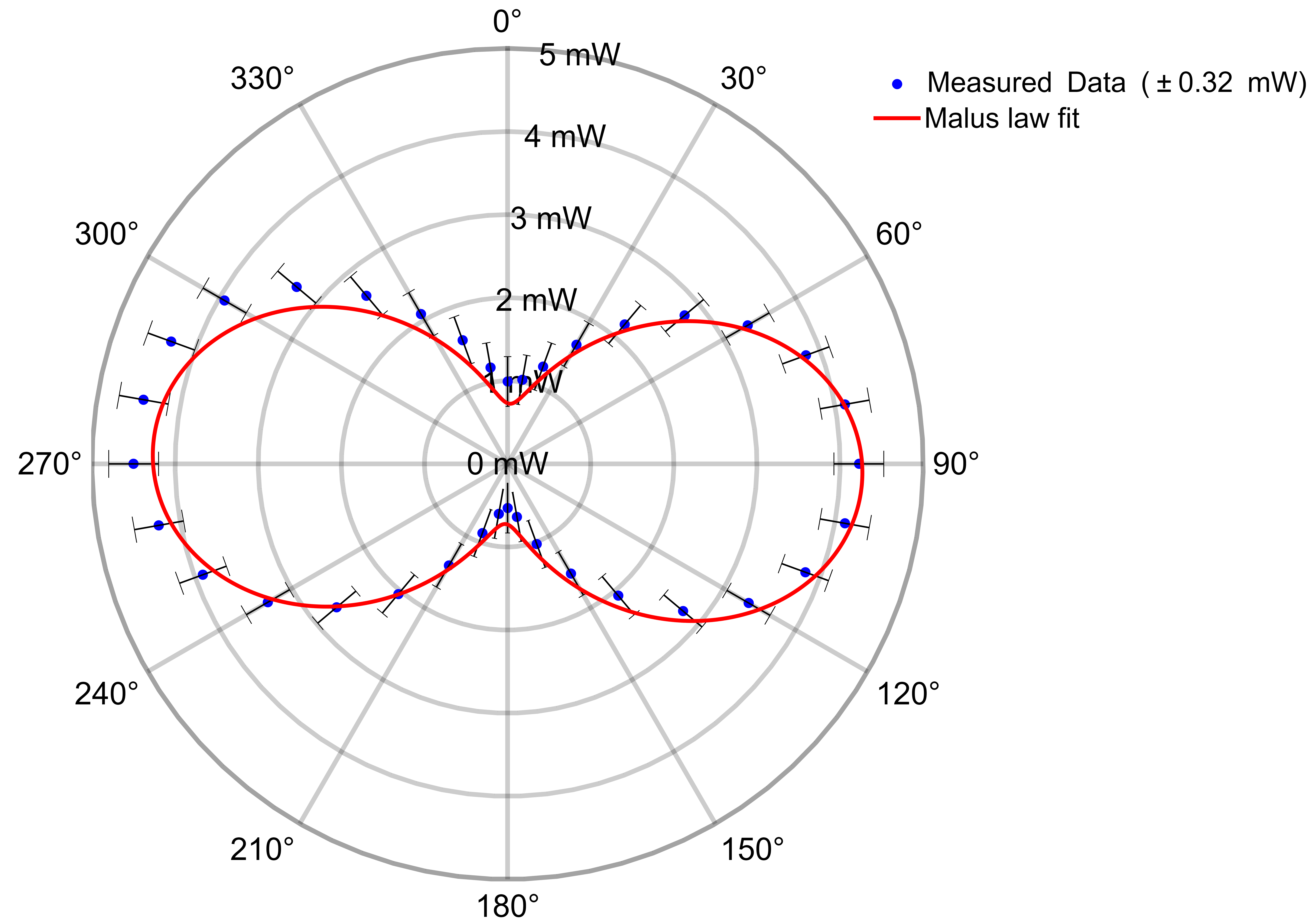}
\caption{Linearly polarized lasing output demonstrating a PER of 10.14 dB \label{Polarization} }
\end{figure}

\subsection{\label{sec:level2}Linearly polarized output}

The polarization properties of the laser output were observed using leakage from the free-space (HR) end of the laser cavity, and the emitted light was found to be linearly polarized with its principal axis aligned along the y-axis (as defined in Fig.~\ref{Microscope}). Figure~\ref{Polarization} shows the angular dependence of the fiber output power, which follows a cosine-squared behavior consistent with Malus’ law, exhibiting a fundamental periodicity of 180\textdegree. The measured intensity displays alternating maxima and minima at approximately 90\textdegree\ intervals, corresponding to orthogonal polarization states.

A polarization extinction ratio (PER) of 10.14 dB was obtained between the orthogonal polarization components. The maximum intensity occurs along the y-axis, while the minimum is observed along the z-axis of the waveguide (see Fig.~\ref{Microscope}).

To investigate the polarization properties of the waveguide and determine whether any appreciable birefringence arises from either the intrinsic material properties or the waveguide geometry (material and form birefringence, respectively), polarization-dependent transmission measurements were performed. For this purpose, a 4.3 mm-long Bragg grating was inscribed within the 5.2 mm-long waveguide and characterized at a resonance wavelength of 2121 nm. This wavelength was deliberately selected outside the principal Ho$^{3+}$ absorption and emission bands to minimize the influence of wavelength-dependent interactions with the active ions and thereby isolate the intrinsic polarization-dependent response of the waveguide.

A supercontinuum laser (LEUKOS Electro VIS470) was used as the probe light source. Its output was linearly polarized using a Glan–Taylor prism, allowing independent measurements of the transverse electric (TE) and transverse magnetic (TM) polarization states. The probe light was coupled into the waveguide using a CaF$_2$ lens (f = 20 mm), while the transmitted light was collected using a ZrF$_4$ optical fiber and analyzed with an optical spectrum analyzer (Yokogawa AQ6375B) at a spectral resolution of 0.05 nm. The measured transmission spectra were normalized against the spectral response of an equivalent waveguide without a grating to compensate for the wavelength-dependent response of the optical system and propagation losses.

\begin{figure}
\includegraphics[trim={0 0 0 0},width=8 cm]{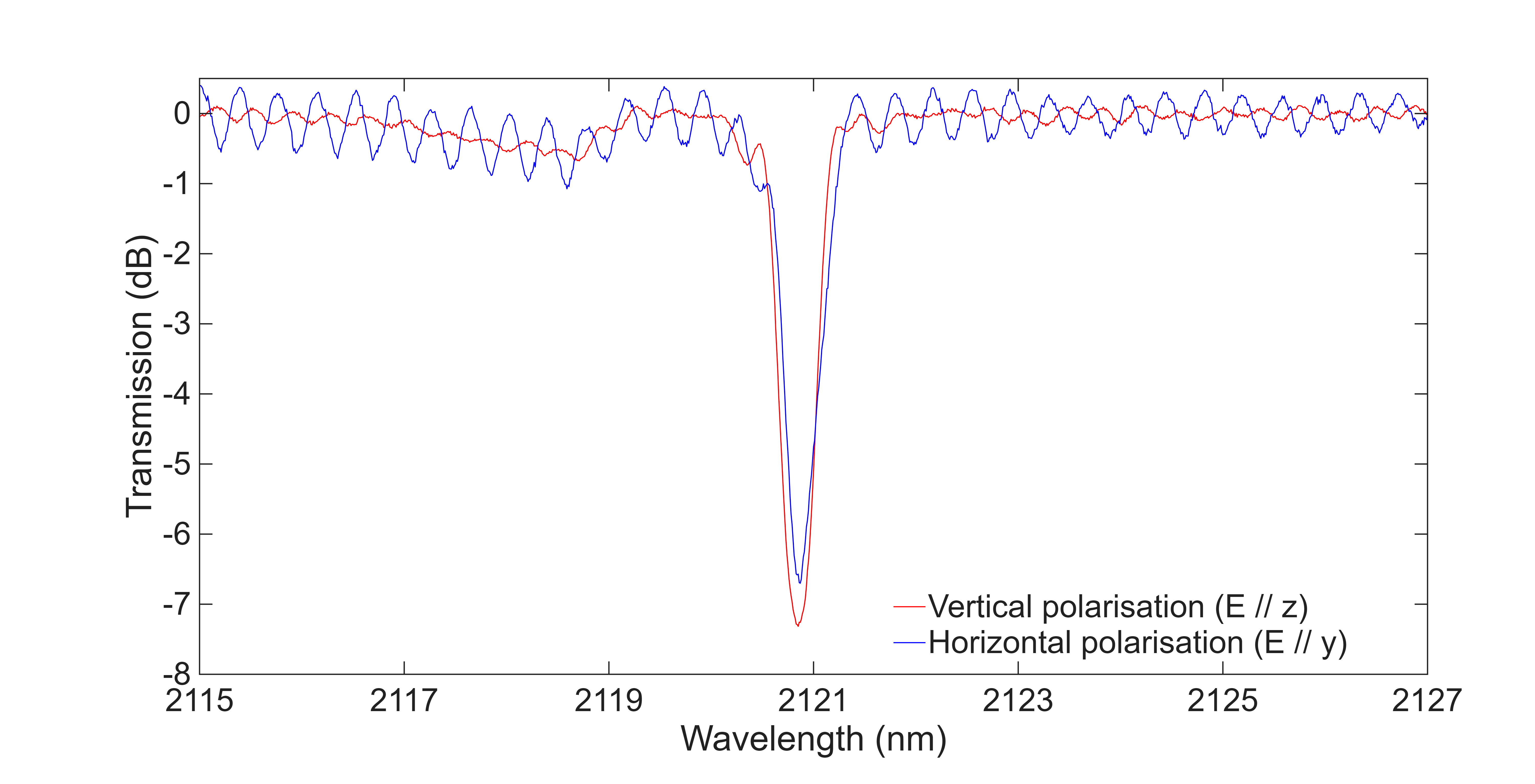}
\caption{Transmission spectra of a waveguide Bragg grating designed for a resonance wavelength outside the water vapor band (2121~nm) for TM ($\mathbf{E} \parallel z$) and TE ($\mathbf{E} \parallel y$) light polarizations. Grating length: 4.3~mm. \label{Birefringence} }
\end{figure}

The resulting transmission spectra for the two orthogonal polarization states are presented in Figure~\ref{Birefringence}. The TE- and TM-polarized spectra exhibit closely matching resonance wavelengths, linewidths (both within the spectral resolution of the analyzer), and transmission-dip depths, with differences remaining within the experimental uncertainty. The small discrepancies observed between the two spectra are attributed primarily to measurement noise, residual artifacts associated with reference normalization, and variations in the coupling efficiency between the optical fiber and waveguide. The close agreement between the TE- and TM-polarized resonance characteristics indicates that no appreciable polarization-dependent response is resolved within the measurement uncertainty, suggesting that the waveguide exhibits negligible effective birefringence at 2121 nm.

\begin{figure}
\includegraphics[trim={0 0 0 0},width=8 cm]{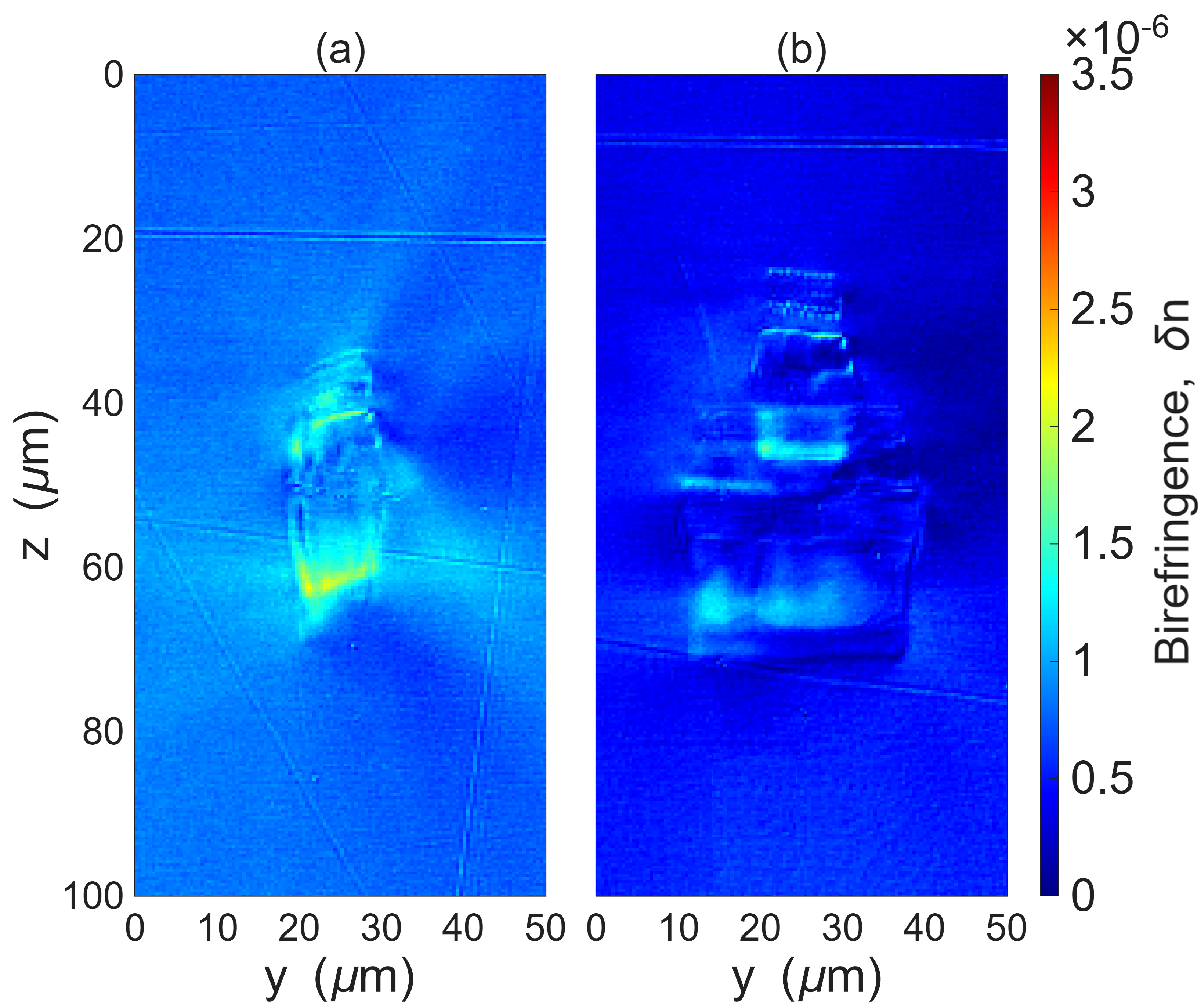}
\caption{Birefringence maps from differential interference contrast microscopy for the structures (a) without and (b) with Bragg gratings \label{Pol_grating} }
\end{figure}

To further investigate the origin and spatial distribution of any residual birefringence, differential interference contrast microscopy was employed to characterize the stress-induced birefringence surrounding the femtosecond-laser-inscribed waveguide and Bragg grating. The measurements were performed using an Olympus BX43F microscope equipped with a two-dimensional birefringence analyzing system (PA-300-MT, Photonic Lattice, Inc.), a 520-nm laser source, and a $\times 50$ objective. Figure~\ref{Pol_grating} shows the spatially resolved birefringence maps calculated from the measured retardance, $\Delta = \delta n \times \ell$, relative to the unmodified glass region. Assuming that the entire sample thickness ($\ell = 5.2$ mm) contributes to the measured retardance, the resulting birefringence distribution closely follows the geometry of the laser-written waveguide. For waveguide-only structures, a maximum birefringence of $\delta n \approx 2.2 \times 10^{-6}$ was estimated. In comparison, for structures incorporating Bragg gratings along approximately 82\% (4.3 mm) of the total waveguide length, the maximum estimated birefringence was slightly lower, at $\delta n \approx 2.0 \times 10^{-6}$. This reduction is tentatively attributed to a possible annealing effect associated with the transverse laser sweep (along the horizontal $y$-axis) during the inscription of the grating elements. Please note that the birefringence values provided should be regarded as an estimate, as the measured retardance may be underestimated due to phase wrapping. Specifically, an undetected additional $2\pi$ phase cycle at the measurement wavelength of 520 nm would correspond to an additional retardance of 520 nm, equivalent to a birefringence contribution of approximately $1.2 \times 10^{-4}$ over the assumed 4.3 mm interaction length. Nevertheless, the measured retardance indicates a relatively weak stress-induced birefringence, which is consistent with the negligible polarization-dependent resonance shifts observed in the Bragg grating measurements.

\begin{figure}
\includegraphics[trim={0 0 0 0},width=8 cm]{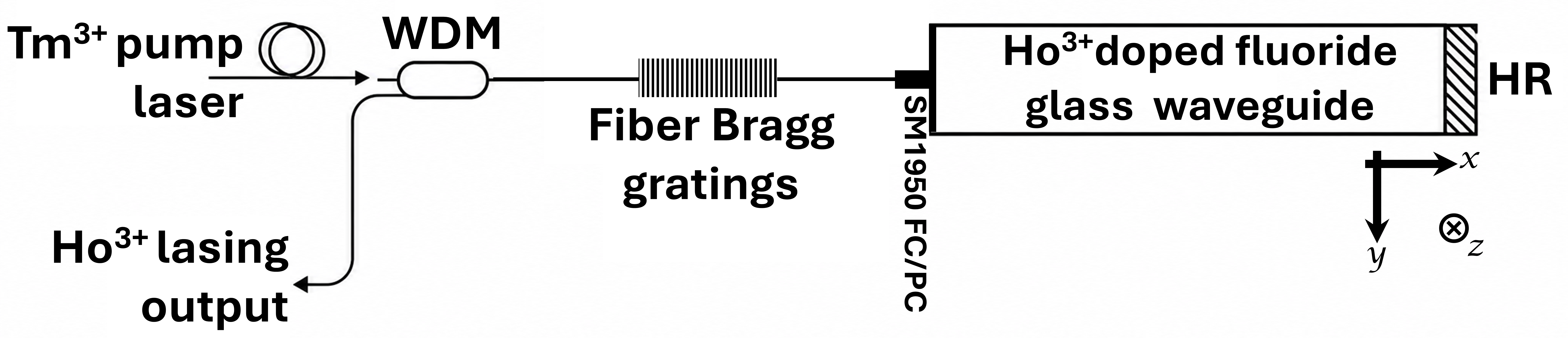}
\caption{Lasing cavity with fiber Bragg grating + HR coating \label{FBG} }
\end{figure}

To investigate the polarization characteristics while lasing with waveguide alone (e.g. without a grating structure) a commercial fiber Bragg grating (FBG) in SM1950 with a reflectivity of 92\% at 2074.48 nm was spliced to the common port of the WDM. This assembly was then butt-coupled to the waveguide, while the opposite waveguide facet was butt-coupled with a broadband high-reflectivity (HR) mirror covering the 2000–2100 nm wavelength range, forming a hybrid fiber/waveguide laser cavity (Fig.~\ref{FBG}).  Lasing was observed at the FBG resonance wavelength of 2074.48 nm. The polarization state of the laser output was examined using an infrared (IR) camera by monitoring the leakage through the HR-coated facet and found to be orthogonal to that observed in waveguides incorporating integrated waveguide Bragg gratings. Specifically, the maximum laser intensity occurred along the z-axis, whereas the minimum intensity was observed along the y-axis of the waveguide (see Fig.~\ref{Microscope}).

\begin{figure}
\includegraphics[trim={0 0 0 0},width=6 cm]{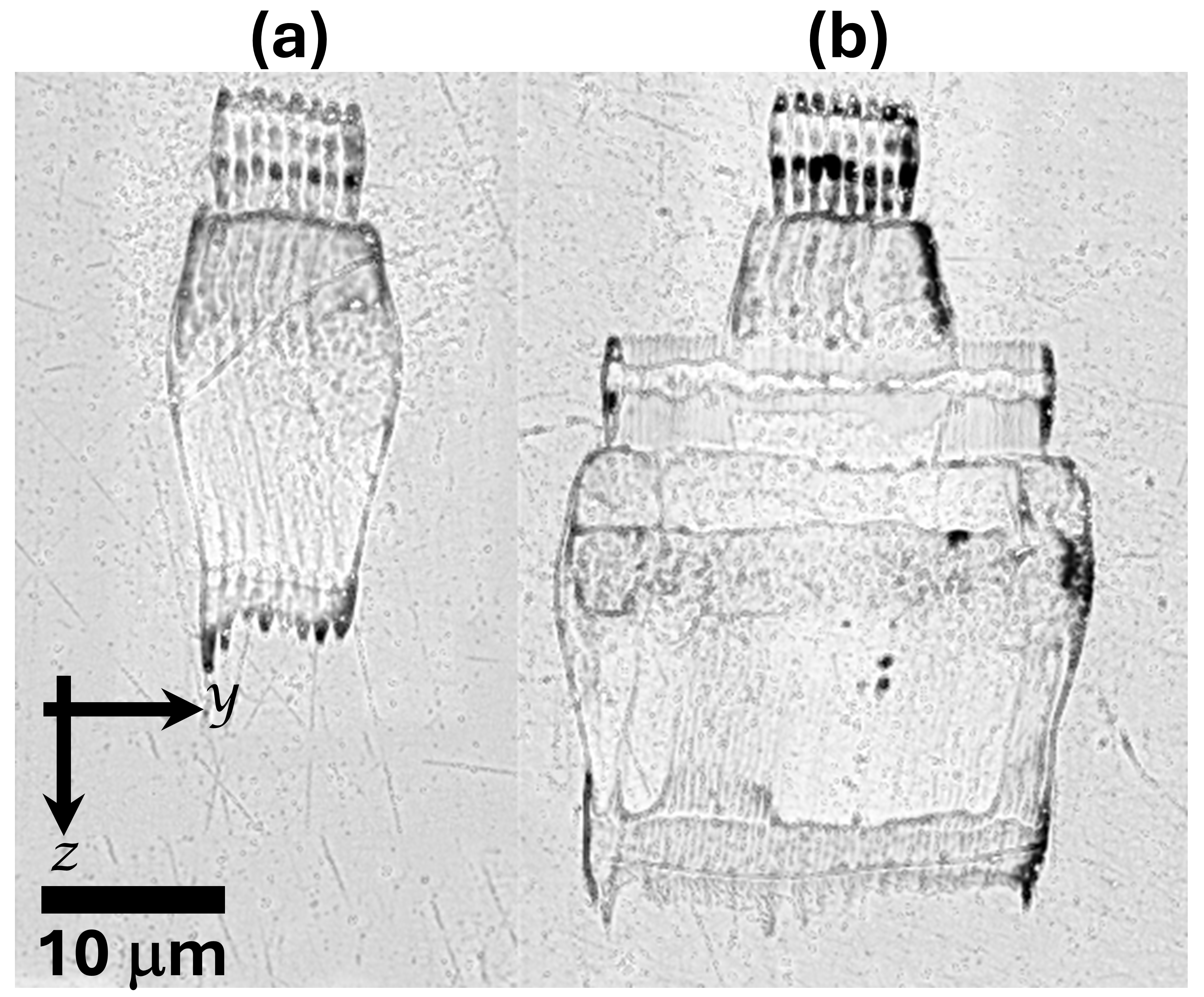}
\caption{BSE images of waveguide (a) without and (b) with Bragg grating structures  \label{BSE}. }
\end{figure}

Polarization-dependent optical guiding has been investigated in a variety of materials, including crystalline and glass-based waveguides. However, to the best of our knowledge, to date no studies have reported polarization-dependent guiding in fluoride glass waveguides. In glass systems, most investigations have focused on fused silica, where Type-I waveguides characterized by a smooth positive refractive index modification generally exhibit negligible polarization dependence. Here we observe fluoride glass waveguides without a super-imposed grating structure (waveguide-only) exhibiting vertically-polarized lasing, and waveguides with a grating structure (WBG) exhibiting horizontally-polarized lasing. Our waveguide-only and waveguide + WBG structures both exhibit (oppositely) polarized lasing output, likely resulting from gain competition between the two polarization modes. For the waveguide-only structure the vertically-polarized lasing suggests increased losses for the horizontally-polarized mode (and vice-versa for the WBG structure). Back-scattered electron microscopy (BSE) images of the waveguide-only and waveguide + WBG transverse cross-sections are found in Fig.~\ref{BSE} (a) and (b) respectively. BSE contrast generally arises from atomic-number (Z) and/or density variations~\cite{Goldstein2018} which typically correlate to variations in the local index of refraction. The observed vertical structures for the waveguide-only system (Fig.~\ref{BSE}(a)) originate from the individual inscription tracks formed by each of the multi-stranded waveguide, forming vertical stripes with a lateral pitch of \textasciitilde{}1 $\mu$m. In the WBG system (Fig.~\ref{BSE}(b)), horizontal structures instead appear in the guiding region, erasing much of the previously observed vertical structures. These sub-wavelength asymmetric features may be a possible origin of the polarization selectivity observed in the two lasing systems.  This interpretation remains tentative and should be regarded as a working hypothesis, as a detailed investigation of the underlying mechanisms responsible for the observed polarization selectivity is ongoing but beyond the scope of the present study. Further investigations will be required to establish the relationship between the observed structural modifications and the polarization-dependent guiding behavior.


\section{\label{sec:level1}Conclusion}
In conclusion, we have demonstrated, for the first time, a fully integrated Ho$^{3+}$-doped fluoride glass waveguide laser incorporating a femtosecond laser-inscribed waveguide and waveguide Bragg grating. The Bragg grating fabrication process was systematically investigated by inscribing gratings with resonance wavelengths spanning 2005--2070~nm and lengths ranging from 5 to 13.5~mm. A pooled standard deviation of 280~pm in Bragg wavelength was obtained, demonstrating high reproducibility and wavelength accuracy across the investigated spectral range. Building on these results, a compact DBR laser cavity was formed using an inscribed Bragg grating and a high-reflectivity mirror. The 12.4~mm-long cavity produced stable single-frequency lasing at 2062.21~nm with a linewidth of 5~MHz. The laser output was linearly polarized and efficiently coupled into an SM1950 single-mode fiber. The combination of reproducible Bragg grating fabrication, monolithic waveguide integration, narrow-linewidth operation, and efficient fiber delivery establishes this alignment-free architecture as a compact and robust platform for integrated single-frequency laser sources operating near 2~$\mu$m.

\begin{acknowledgments}
The authors acknowledge the support received from the Commonwealth of Australia through an Australian Research Council (ARC) Discovery Grant No. DP220102516 and Centre of Excellence Grant Nos. CE170100004 and CE230100016. This research was supported by the Australian Research Council Centre of Excellence in Optical Microcombs for Breakthrough Science (project number CE230100006) and funded by the Australian Government. TTF and DL acknowledge support received from Electro Optic Systems Pty. Limited.  This work was performed in part at the Opto Fab node of the Australian National Fabrication Facility supported by the Commonwealth and SA State Government. The authors acknowledge the use of facilities supported by Microscopy Australia at the Electron Microscope Unit within the Mark Wainwright Analytical Center at UNSW Sydney.
\end{acknowledgments}

\section*{Data Availability Statement}

The data that support the findings of this study are available from the corresponding author upon reasonable request.

\nocite{*}
\bibliography{References}

\end{document}